\documentclass[epj]{svjour}
\usepackage{graphics}
\usepackage{latexsym, amssymb, amsmath}
\usepackage[dvips]{graphicx}
\usepackage{epsfig,subfigure}
\usepackage{picinpar}
\usepackage{eufrak}
\usepackage{fancyhdr}

\begin{document}

\title{Photoconductance of a one-dimensional quantum dot}
\authorrunning{M. Vicari {\em et al.}} 
\titlerunning{Photoconductance of a one-dimensional quantum dot}
\author{M. Vicari, A. Braggio, E. Galleani d'Agliano, M. Sassetti}

\institute{Dipartimento di Fisica, INFM, Via Dodecaneso 33, I-16146 Genova, 
Italy}

\date{Received: 22 October 2001  / Revised version: date}

\abstract{The ac-transport properties of a one-dimensional quantum dot
  with non-Fermi liquid correlations are investigated. It is found
  that the linear photoconductance is drastically influenced by
  the interaction. While for weak interaction it shows peak-like
  resonances, in the strong interaction regime it assumes a step-like
  behavior.  In both cases the photo-transport provides precise
  informations about the quantized plasmon modes in the dot.
  Temperature and voltage dependences of the sideband peaks are
  treated in detail.  Characteristic Luttinger liquid power laws are
  found.}


\maketitle

\section{Introduction}
\label{sec:1}

In the last years, theoretical and experimental efforts have focused
on the analysis and control of transport in low dimensional quantum
systems. The experimental realization of one-dimensional (1D) quantum
wires has opened new possibilities to investigate the influences of
interactions and impurities on electron transport in the quantum
regime \cite{tar95,yac96,yac97}. Decreasing the electron density in
the cleaved-edge-over\-growth GaAs/AlGaAs quan\-tum wire, it was
possible to reach the region of Cou\-lomb blockade where transport
occurs by transferring exactly one electron through a {\em 1D quantum
  island} created by random impurities \cite{aus}.  The temperature
dependence of the conductance Coulomb peak exhibited, for the first
time, a clear evidence of a non-Fermi liquid behavior.  Similar
behaviors were also indicated by recent results of Raman scattering,
which strongly suggests that semiconductor quantum wires are very
probably dominated by non-Fermi liquid excitations \cite{sk98}.

The existing broad theoretical understanding of 1D interacting electrons 
\cite{lutt,hal,Voit} provided in the past quantitative results for
{\em dc-transport} in the presence of correlations and impurities.  1D
quantum dots embedded in a Luttinger liquid were subject of several
theoretical investigations
\cite{KaneFis,FurNag,SasNapWei,MauGia,Furusaki98,braggio99}.
Microscopic charging effects and transport properties in the
sequential and cotunneling regimes were discussed. Important results,
connected with finite range interactions and spin dynamics were also obtained
\cite{Tob,br00}.

Recently, frequency and time dependent effects have been investigated
intensively.  When an ac-field is applied, the conduction properties
of the system can be stron\-gly modified.  Indeed an ac-polarization
enables to conduct through new photon-assisted channels due to
inelastic scattering of electrons and photons.  The effect of the
above photon assisted tunneling (PAT) becomes crucial in the presence
of a quantum dot. Here, the electrons overcome the Coulomb blockade
regime by absorbing photons from the external field.  These additional
transport channels enrich the conductance structure and give precise
information about ground state and excited states in the
dot.\\
$^{}$\hspace{0.6cm}PAT is a topic studied since about forty years. A
pa\-ra\-dig\-ma\-tic result was obtained by Tien and Gordon \cite{TG}
for a superconducting tunnel junction.  Resonant tunneling in a
periodic time-dependent external field through a double barrier was
investigated within the framework of classical perturbation theory in
order to examine the effects due to the interplay between the lifetime
of an electron trapped in the well and the microwave period
\cite{Sok}.  PAT has been investigated for 2D quantum dots in the
region of charging effects and intra-dot transitions both in
theoretical \cite{Brud,BruBrud1} and experimental works
\cite{Kou,Kou1,Bli,Ooster,Oost98}. In the mean-field approximation,
self-consistent renormalization of the driving field has been also
studied \cite{Ped}. More recently PAT has been analyzed to determine
the complex microwave photoconductance of a single quantum dot
\cite{Qin}, but a great interest has also focused on systems formed by
weakly coupled quantum dots \cite{Fuj}.  In spite of this broad
researches, mainly devoted to 2D systems, there is still a lack of
investigations related to 1D quantum dots in the presence of
ac-fields.  Here, a microspic treatment of the interplay between
correlation and ac-fields is crucial and can yield new information on
the physics of non-Fermi liquids.  In the past, ac-transport in a
Luttinger liquid through a {\em single} barrier has been considered,
including higher harmonics generation, frequency scaling properties
and local fields corrections
\cite{sas96,fech99,cun99}.\\
$^{}$ \hspace{0.3cm} In this paper we present results of time
dependent transport through a {\em double} barrier.  We evaluate the
pho\-to-con\-ductance of a 1D quantum dot as a response to a
monochromatic signal in the sequential tunneling regime.  The electron
interactions are treated within the framework of the Luttinger liquid.
We show that the ac-conduction features are strongly affected by the
strength of the interaction. The latter gives rise to two particular
regimes of {\it strong} and {\it weak} interaction. The difference
between the two appears in the temperature and/or gate voltage
dependence of the photoconductance. By varying the frequency we
identify sideband peaks related to the ground and the excited states.
The latter are due to the collective quantized plasmons in the dot.
Non-analytic power laws of the peaks are also predicted and their
dependence on the interaction is considered.

The paper is organized as follows. In Section 2 we describe the model.
The 1D quantum dot is confined inside the single channel quantum wire
by two high localized scattering barriers. Transport is tuned by 
gate and  source-drain voltages.  In Section 3 we present the solution
for ac-transport in the regime of high frequencies. The current is
described by a master equation \cite{Bee,in92,braggio99}, generalized
to the case of time-dependent polarizations.  Section 4
is devoted to the discussion of the behavior of the linear
photoconductance. In Section 5 we draw conclusions and  discuss also the the
consistency of the approximations.

\section{The model}
\label{sec:2}
We consider 1D spinless interacting electrons treated via the
bosonization technique \cite{Voit}. The quantum dot is described by two
symmetric delta-like barriers $V_{\rm B}\,\delta(x-x_{i})$ at $x_{i}$ ($i=1,2$)
where $x_{1}<x_{2}$. External dc- and ac-potentials are coupled to the
dot in order to induce transport. The Hamiltonian is 
\begin{equation}
H=H_{0}+H_{\rm{B}}+H_{\rm{V}}.
\end{equation}
The first term describes the collective low-energy charge density modes 
of the 1D electron gas ($\hbar=1$) \cite{lutt,Voit} 
\begin{equation}
H_0=\frac{v_{\rm{F}}}{2}\int^{+\infty}_{-\infty}dx 
\Big[ \Pi^2 (x) +\frac{1}{g^2}\big(\partial_{x}\Theta(x)\big)^2\Big]\,.
\label{diagonal}
\end{equation}
Their quantization is
realized by the field operator $\Theta (x)$ and its conjugate $\Pi(x)$.
The former represents the long-wavelength part of the electron number density  
\begin{equation}
\rho(x)=\rho_{0}+\sqrt{\frac{1}{\pi}}\partial_{x}\Theta(x)\,,
\label{curop}
\end{equation}
with the mean electron density $\rho_{0}=k_{\rm F}/\pi$ ($k_{\rm F}$
Fermi wave number).  The velocity of
the charge modes is renormalized with respect to the Fermi velocity
$v_{\rm F}$, because of the electron interaction.  It is $v=v_{\rm
  F}/g$, with the interaction strength
\begin{equation} 
\frac{1}{g}=\sqrt{1+\frac{V(q\to 0)}{\pi v_{\rm{F}}}}.
\end{equation}
Here, $V(q)$ is the Fourier transform of the 3D Coulomb interaction 
projected along the wire \cite{sk98}.

The contribution of the two localized impurities involves $2k_{\rm
  F}$-backscattering interference between left and right moving
electrons.  It can be written in the bosonized form as
\cite{FurNag,SasNapWei}
\begin{equation}\label{sinpot}
H_{\rm{B}}=U_{\rm{B}}\cos(\pi N_{+})\cos[\pi(n_{0}+N_{-})]\,,
\end{equation}
where $U_{\rm B}\equiv\rho_0 V_{\rm B}$, and
$N_{\pm}=[\Theta(x_{2})\pm\Theta(x_{1})]/{\sqrt{\pi}}$.
The quantity  $N_{+}/2$, is associated with the unbalanced particles 
between left and right leads, while $N_{-}$ represents the fluctuations of 
the particle number in the dot with respect to the  mean electron number 
$n_{0}=\rho_{0}d$; ($d=x_{2}-x_{1}$).  

The external fields consist of a dc-source-drain voltage $V$ that
drops symmetrically at the barrier, and a gate voltage, renormalized
by the ratio between gate and total capacitance, with a
monochromatic component
$V_{\rm{g}}+{V}_{\rm{ac}}\cos(\omega t)$.  The coupling is then
($e>0$)
\begin{equation}
H_{\rm{V}}=-e\left[ \frac{V}{2}N_{+}+\left(V_{\rm{g}}+{V}_{\rm{ac}}
\cos(\omega t)\right)N_{-}\right]\,,
\label{Hv}
\end{equation} 

In the  presence of ac-fields the {\em particle} current depends on the spatial
position along the wire \cite{sassetti96}.
In the stationary limit ($t\to\infty$) it 
will be given by a superposition of all harmonics of the monochromatic signal
\begin{equation}
\label{corrente}
I_{\rm st} (x,t)
=\sum^{+\infty}_{l=-\infty}I^{(l)}(x)\,e^{-il\omega t}\,.
\end{equation}
Moreover, there will be {\em displacement} contributions that
eventually will renormalize the total current \cite{Ped}.

In the following we will focus only on the steady component ($l=0$) of
the total current. This term is dominated by the particle current and
allows us to neglect displacement contributions.  It is space
independent and can then be evaluated from the time derivative of the
number of transferred particles through the dot
\begin{equation}\label{corr}
I(t)=\frac{e}{2}<\dot{N}_{+}(t)>\,.
\end{equation}
The bracket $<...>$ includes a thermal average over the collective
charge density modes away from the barriers and a statistical average
over the final states of the reduced density matrix for the degrees of
freedom at $x=x_{1},x_{2}$.  The former play the role of a thermal
bath. They act as a dissipative source for the degrees of freedom at
the positions of the barriers and they can be exactly traced out. 
The explicit contribution of their integration
consists of two terms \cite{sassetti96} . 

The first term is frequency independent and
it is responsible to suppress changes in the particle number on the
island, with respect to the equilibrium value. The corresponding energy 
is $E_{\rm c}N_{-}^2$ with the charge addition energy  
\begin{equation}
E_{\rm c}=\frac{\pi v_{\rm F}}{2d g^2}\,.
\end{equation}
Without interaction ($g=1$), the addition energy is still finite 
$E_{\rm c}=\pi v_{\rm F}/2d=E_{\rm P}$, due to 
the Pauli principle, and the discreteness of the dot levels. On the other hand,
for strong Coulomb interaction $E_{\rm c}\propto V(q\to 0)\gg E_{\rm P}$.

The second term is frequency dependent and represents the dynamical effects
of the external leads and of the correlated excited states in the dot.
Its influence is described by the spectral density \cite{FurNag,SasNapWei}
\begin{equation}
J(\omega)=\frac{ \omega}{g}\Big[1+\varepsilon\sum_{m=1}^\infty 
\delta(\omega -m\varepsilon)\Big]
\equiv J_{\rm leads}(\omega) + J_{\rm dot}(\omega)\,.
\label{spectral1}
\end{equation}
The first term is due to the leads, while the second
reflects the discretization energy $\varepsilon$ of the charge modes
in the quantum dot
\begin{equation}
\label{dotenergy}
\varepsilon=\frac{\pi v_{\rm F}}{d g}\equiv 2gE_{\rm{c}}\,. 
\end{equation} 
\section{AC-transport}
\label{sec:3}
For obtaining the time dependent current we have to consider the
dynamics of the $N_{\pm}$-variables under the influences of the
external fields (\ref{Hv}), the dissipation (\ref{spectral1}) and the
2D periodic potential (\ref{sinpot}).  For high barriers
$U_{\rm{B}}\gg E_{\rm{c}}$, the dynamics is dominated by tunneling
events between nearest-neighbored minima, with amplitude $\Delta$ that
is related to $U_{\rm{B}}$ via WKB-approximation.
In the following, we will consider the contribution of uncorrelated
hops off and onto the island (sequential tunneling processes) with the
temperature smaller then the discretization energy $\varepsilon$.

The main quantities, in this case, are the tunneling rates.
We identify with $\Gamma(U_{\nu,s}(n),t)$ the rate of hops throu\-gh the
right ($\nu={\rm r}$) or left ($\nu ={\rm l}$) barrier, in the forward
($s={\rm f}$) or backward ($s={\rm b}$) direction, respectively
with initially $n\equiv n_0+N_{-}$ number of electrons in the dot.
To the lowest order in the tunneling matrix element
$\Delta$, one has \cite{fech99}
\begin{equation}
\Gamma(U_{\nu,s}(n),t) = \int_{0}^{t} \!d t'  \kappa_{\nu,s}(n,t,t')\,,
\label{Gamma}
\end{equation}
where
\begin{eqnarray}
\kappa_{\nu,s} (n,t,t') &=&\frac{\Delta^2}{4}
\exp\Big[-W(t-t')+iU_{\nu,s}(n)(t-t')\nonumber\\
&&-i e
\chi_{\nu,s}V_{\rm{ac}}\int_{t'}^{t} \!d\tau\,
\cos(\omega \tau)\Big]+c.c.\,.
\label{kappa}
\end{eqnarray}
Here $\chi_{\nu,{\rm f}}=-\chi_{\nu,{\rm b}}=\pm 1$ for $\nu={\rm r,l}$. 
The function $W(t)$ describes the influence of bulk modes, it is
directly connected to the spectral density (\ref{spectral1}) by \cite{fech99} 
($\beta=1/k_{\rm{B}}T$),
\begin{equation}
W(t)=\int^{\infty}_{0}d\omega \frac{J(\omega)}{\omega^2}\Big 
[(1-\cos\omega t)\coth\frac{\beta\omega}{2}+i\sin\omega t\Big ]\,.
\label{w}\end{equation}
The energies $U_{\nu,s}(n)$ are the differences between initial 
and final addition and electrostatic energies
associated with different types of hops ($\nu,s)$
\begin{eqnarray}
U_{\nu,s}(n)&=&2E_{\rm{c}}\left[\chi_{\nu,s}
(n-n_{\rm g}) -\frac{1}{2}\right]-\eta_{s}\frac{eV}{2}\,,\nonumber\\
n_{\rm g}&\equiv& n_0+\frac{e V_{\rm g}}{2E_{\rm c}}\,.
\label{U}
\end{eqnarray}
The reference particle number $n_{\rm g}$
is defined by the gate voltage, and $\eta_{s}=\pm 1$ for
$s={\rm f,b}$. The lack of time-invariance due to the ac-field is
reflected in the double-time dependence of the kernel $\kappa(t,t')$.  
In the absence of the ac-field, $\kappa(t,t')$ is
translational invariant and the corresponding dc-rate is
the $t\to \infty$ limit of (\ref{Gamma}) for $V_{\rm ac}=0$
\begin{equation}\label{GDC}
\Gamma_{\rm dc}(U_{\nu,s}(n))=\frac{\Delta^2}{4}\int^{+\infty}_{-\infty}dt\; 
e^{-W(t)}e^{iU_{\nu,s}(n)t}\,.
\end{equation}
This integral can be explicitely evaluated at low temperatures, 
$k_{\rm B}T\ll\varepsilon$, \cite{braggio99}
\begin{equation}
\Gamma_{\rm {dc}}(U)=\sum_{p=-\infty}^{\infty}
 w_p(\varepsilon){\gamma}^{(2g)}(U-p\varepsilon)
\label{series}
\end{equation}
where $\gamma^{(g)}(U)$ is the tunneling rate of a single barrier
\begin{eqnarray}
\gamma^{(g)}(U)&=&\frac{\Delta^2}{4\omega_{\rm c}}
\frac{e^{\beta U/2}\,e^{-\vert U\vert/\omega_{c}}}{\boldsymbol{\Gamma}(2/g)}
\Big(\frac{\beta\omega_{\rm c}}{2\pi}\Big)^{1-2/g}\nonumber\\
&&\times\left|\boldsymbol{\Gamma}
\Big(\frac{1}{g} +
{\rm i}\frac{\beta U}{2\pi}\Big)\right|^2 \,,
\label{gammaom}
\end{eqnarray}
with frequency cutoff $\omega_{\rm c}$ and $\boldsymbol{\Gamma}(z)$
the Gamma function.  The weights $w_p(\varepsilon)$, at discrete 
energies $\varepsilon$, are for $k_{\rm B}T\ll\varepsilon$
\begin{equation}
w_p(\varepsilon)=\theta(p)
\left(1-e^{-\varepsilon/\omega_c}\right)^{1/g}
\frac{\boldsymbol{\Gamma}(1/g+p)}{p!\,\boldsymbol{\Gamma}(1/g)}
\,e^{-p\varepsilon/\omega_{\rm c}}\,.
\end{equation}

In the presence of the ac-field, the kernel $\kappa(\tau,\tau-\tau')$ depends
periodically on $\tau$ with a period ${\cal T}=2\pi/\omega$.  This allows
to write the stationary ($t\to\infty$) rate
(\ref{Gamma}) in terms of principal and higher harmonics of the
monochromatic field
\begin{equation}
\label{statgam}
\Gamma_{\rm st} (U_{\nu,s}(n),t)
=\sum^{+\infty}_{l=-\infty}\Gamma^{(l)}(U_{\nu,s}(n))\,e^{-il\omega t}\,.
\end{equation}

In the sequential tunneling regime the expression for the
corresponding current is characterized by the sum over all possible
contributions stemming from different electron numbers $n$ in the
island, weighted by the occupation probability $p(n,t)$
\begin{equation}\label{I}
I(t)= \frac{e}{2}\sum^{+\infty}_{n=-\infty}\int^{t}_{0} dt'p(n,t')
[\kappa_{{\rm b}}(n,t,t')-\kappa_{{\rm f}}(n,t,t')]\,,
\end{equation}
where $\kappa_{s}= \kappa_{{\rm r},s} +\kappa_{{\rm l},s}$.
The probability $p(n,t)$ is determined by the non-Markovian 
master equation \cite{GrifSas}
\begin{eqnarray}\label{GME}
&&\hskip-0.5cm\dot{p}(n,t)=-\int^{t}_{0} dt'\Bigl\{
p(n,t')\left[\kappa_{{\rm f}}(n,t,t')+\kappa_{{\rm b}}(n,t,t')\right]
\nonumber\\&&\nonumber\\&&\hskip-0.5cm
-p(n+1,t')\left[\kappa_{{\rm l,\rm b}}(n+1,t,t')+
\kappa_{{\rm r,\rm f}}(n+1,t,t')\right]
\nonumber\\&&\nonumber\\
&&\hskip-0.5cm-p(n-1,t')\left[\kappa_{{\rm r,\rm b}}(n-1,t,t')+
\kappa_{{\rm l,\rm f}}(n-1,t,t')\right]\Bigr\}\,.
\end{eqnarray}
Because of the ac-fields, the asymptotic solution of (\ref{GME})
will consist of a 
superposition of all harmonics
\begin{equation}\label{statprob}
  p_{\rm st}(n,t) = \sum^{+\infty}_{l=-\infty}p^{(l)}(n)\,e^{-il\omega t}\,.
\end{equation} 
Inserting this expansion in the above master equation one obtains an
infinite set of linear equations in which the $p^{(l)}(n)$ are coupled
with each other via the $\kappa_{\nu,s}$ kernels. These equations can
be solved recursively. For details see the similar discussion 
performed for a driven dissipative many states system \cite{GrifSas}.

Here we are interested in the limit of high frequencies.  For $\omega\gg
\{\Gamma^{(0)}(U_{\nu,s}(n)),\Delta,eV\}$, the ac-field oscillates too
fast to be able to catch the details of the dynamics within one period
${\cal T}$.  This corresponds to the non-adiabatic re\-gi\-me
where each electron experiences many cycles of the ac-field during its
presence inside the dot. In this limit one can approximate the 
kernels $\kappa_{{\nu,\rm s}}(m,t,t')$ in (\ref{GME}) by a time average
\cite{GrifSas}
\begin{eqnarray}
&&\kappa_{{\nu,\rm s}}(m,t,t')\approx\kappa^{(0)}_{{\nu,\rm s}}(m,t-t')=
\nonumber\\
&&~~~=\frac{1}{{\cal T}}\int\limits_{0}^{{\cal T}}
d\tau\kappa_{{\nu,\rm s}}(m,\tau,\tau-(t-t'))\,,
\end{eqnarray}
reducing the master equation (\ref{GME}) to a convolutive form.  This
implies a stationary solution without periodic oscillations as in
(\ref{statprob}). It can be written in terms of the lowest Fourier
component $p^{(0)}(n)$ that satisfies detailed balance 
\begin{eqnarray}
\label {p0}
&&\hskip-0.3cm p^{(0)}(n)[ {\Gamma}^{(0)}(U_{{\rm l,\rm f}}(n)) 
+ {\Gamma}^{(0)}(U_{{\rm r,\rm b}}(n))]=
\\&&\nonumber\\&&\hskip-0.5cm 
=p^{(0)}(n+1)[{\Gamma}^{(0)}
(U_{{\rm l,\rm b}}(n+1)) + 
{\Gamma}^{(0)}(U_{{\rm r,\rm f}}(n+1))]\,,\nonumber   
\end{eqnarray}
with the normalization constraint $\sum^{+\infty}_{n=-\infty}
p^{(0)}(n) =1$. 
The average rate
\begin{equation}
\label{ratemediato}
{\Gamma}^{(0)}(U_{{\nu,\rm s}}(m))=\int\limits_{0}^{\infty}
d\tau\kappa^{(0)}_{{\nu,\rm s}}(m,\tau)\,,
\end{equation}
represents the Fourier component corresponding to $l=0$ of the
asymptotic rate (\ref{statgam}).  From the periodicity of the external
field it can be written as a superposition of the dc-rate
$\Gamma_{\rm{dc}}$ evaluated at integer multiples of the frequency
$\omega$ \cite{cun99}.
\begin{equation}\label{GAC}
{\Gamma}^{(0)}(U_{\nu,s}(n))=\sum^{+\infty}_
{k=-\infty}J^2_{k}\left(\frac{eV_{\rm{ac}}}{\omega}\right)\Gamma_{\rm{dc}}
(U_{\nu,s}(n)+k\omega)\,,
\end{equation}                      
here $J_{k}(x)$ is the Bessel function of order $k$.

Performing equivalent high
frequency approximations on the kernels (\ref{I}), one obtains 
the asymptotic and steady component ($l=0$) of the current (\ref{corrente}) 
\begin{eqnarray}\label{corprob}
 I^{(0)} &=&\frac{e}{2}  \sum^{+\infty}_{n=-\infty} p^{(0)}(n)
\left[{\Gamma}^{(0)}(U_{\rm l,\rm b}(n))+{\Gamma}^{(0)}(U_{\rm r,\rm b}(n))
\right.\nonumber\\
&&\left.-{\Gamma}^{(0)}(U_{\rm l,\rm f}(n))-{\Gamma}^{(0)}
(U_{\rm r,\rm f}(n))\right]\,.
\end{eqnarray}

In the following we will consider linear transport, $V\!\to \!0$, 
in Coulomb blockade regime $E_{\rm{c}}\gg
\varepsilon,k_{\rm{B}}T,\omega$, where only two channels (say $n$ and
$n+1$) contribute appreciably to a given resonance conductance peak.
The energies $U_{\nu,s}(n)$ and $U_{\nu,s}(n+1)$ given by (\ref{U})
are proportional to the gate energy $\mu_{\rm{g}}=e(V_{\rm{g}}-V^{\rm 
res}_{\rm{g}})$ that measures the shift with respect to the gate
resonance value $eV^{\rm res}_{\rm{g}}=E_{\rm{c}}[2(n-n_{0})+1]$.
The photoconductance can then be written in terms of a single
rate ${\Gamma}^{(0)}(\mu_{\rm g})$
\begin{equation}\label{lincond}
G(\mu_{\rm g})=\frac{e^2}{2}\cdot\frac{\sum_{l=\pm}{\Gamma}^{(0)}(l\mu_{g})
{\Gamma}'(-l\mu_{\rm g})}{\sum_{l=\pm}{\Gamma}^{(0)}
(l\mu_{\rm{g}})},
\end{equation}
where $\Gamma'(x)\equiv d\Gamma^{(0)}(x)/dx$.
In the absence of the ac-field this reduces to the dc-case of
the linear conductance \cite{braggio99}
\begin{equation}\label{dclincond}
G_{\rm dc}(\mu_{\rm g})=\frac{e^2}{4}\cdot
\frac{e^{-\beta\mu_{\rm g}/2}\Gamma_{\rm dc}(\mu_{g})}
{k_{\rm B}T\cosh(\beta\mu_{\rm g}/2)}\,.
\end{equation}

\section{Photoconductance}
\label{sec:4}
In this section we analyze the behavior of the linear
photoconductance in the discretization regime $k_{\rm B} T\ll\varepsilon$. 
In order to do this it is useful to recall the
characteristic energy scales for the
transport. The quantum dot is characterized by a ground state
energy of $n$ charge $E_0(n)=E_{\rm c}(n-n_{\rm g})^2$. The energy
differences of the many body states of $n+1$ and $n$ electrons are
\begin{equation}
\mu(n,l)=2E_{\rm c}\left(n-n_{\rm g}+\frac{1}{2}\right)+l\varepsilon\,.
\end{equation}
Positive or negative integers $l$ denote the differences of the number
of charge excitation quanta with the discretization (\ref{dotenergy}).
They do not change the number of particle in the dot.  The quantity
$\mu(n,l)$ plays the role of chemical potential of the dot and defines
the transport region.

\begin{figure}
\resizebox{0.48\textwidth}{!}{%
  \includegraphics{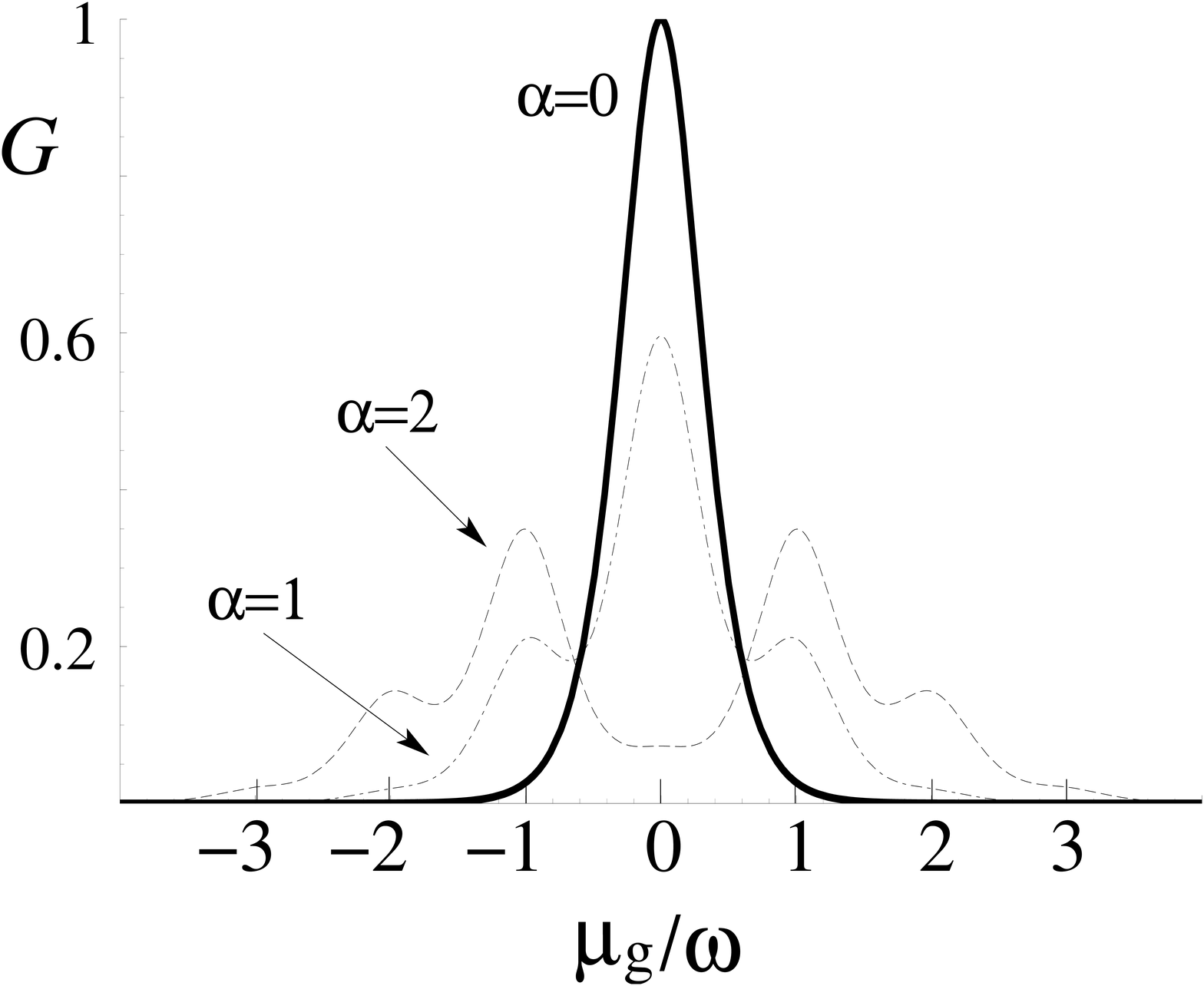}
}
\caption{Linear conductance, 
normalized to the maximum of the dc-resonance
 (thick line), in the non-interacting case $g=1$, 
as a function of the gate energy $\mu_{g}/\omega$.  
Different curves correspond to different values of 
$\alpha=eV_{\rm{ac}}/\omega$, for $\varepsilon/\omega=3.5$, 
$k_{\rm B}T=\omega/5$, $\omega/\omega_{c}=10^{-4}$.} 
\label{fig1}       
\end{figure}

We start now to discuss the simpler case of large discretization
energy $\varepsilon>\omega$.  Here we expect that the collective
charge excitation quanta cannot be excited by the photons, and only
the ground state chemical potential $\mu(n,l=0)$ will be involved.
This is shown in Fig. \ref{fig1}. The conductance is
plotted as a function of the gate energy $\mu_{\rm{g}}$ for different
values of the parameter $\alpha=eV_{\rm ac}/\omega$ and $g=1$.
Temperature is low enough ($k_{\rm{B}}T<\omega$) in order to observe
the discreteness of the photon energy. The  curve for $\alpha=0$
corresponds to the dc-case (\ref{dclincond}). Here the conduction is
present only when $\mu(n,l=0)$ lines up with the Fermi levels of the
leads.  The width of the resonance is due to the finite temperature.
Away from the voltage range characterizing the resonance peak, the
system is affected by Coulomb blockade and does not conduct within the
sequential tunneling regime. The transport properties radically change
when we turn on an ac-field. By modulating the gate polarization
$\alpha$ the conduction electrons exchange photons with the external
field.  The corresponding inelastic electron-photon scattering results
in the splitting of the dc-resonance into a series of principal
resonance sidebands.  They show up at integer value of the frequency
$\mu_{\rm g}=m\omega$ with $ m=\pm 1,\pm2\dots$.

\begin{figure}
\resizebox{0.48\textwidth}{!}{%
  \includegraphics{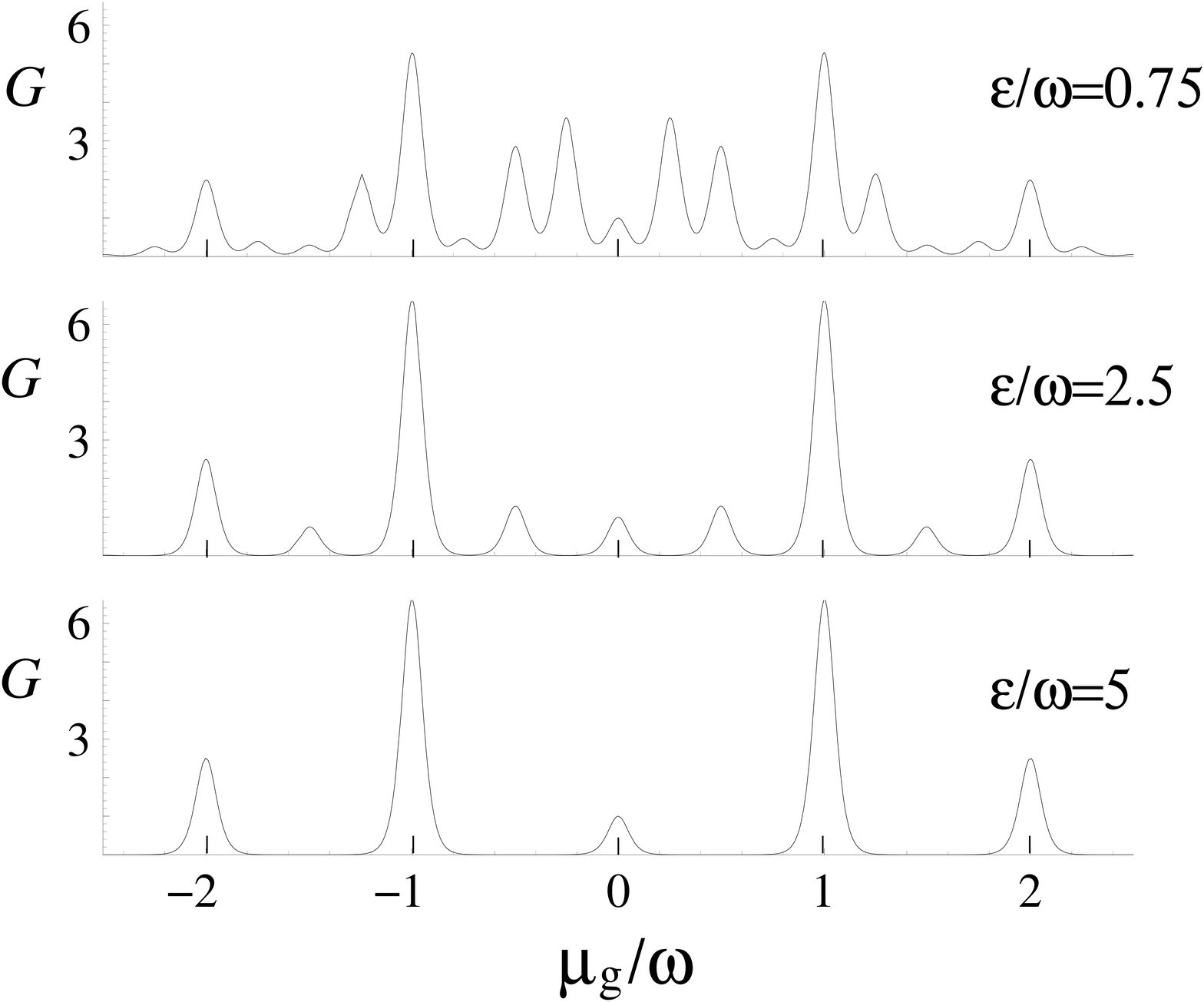}
}
\caption{Linear conductance, normalized to the value of the central resonance 
at $\mu_{g}=0$, as a function of the 
gate energy $\mu_{g}/\omega$, 
for  $\alpha=2$, $g=1$, 
$k_{\rm B}T=\omega/30$, $\omega/\omega_{c}=10^{-4}$ and 
different values of $\varepsilon/\omega$.} 
\label{fig2}       
\end{figure}

Increasing the frequency, transitions to excited charge mode states
can also be supported.  The corresponding structures in the
conductance are shown in Fig. \ref{fig2}. This {\it multi-level
  regime} is characterized by resonance peaks due to the interplay of
discrete charge quanta, with inelastic electron-photon scattering.
Their energy positions are situated at
$\mu_{\rm{g}}=m\omega+l\varepsilon$, with $m,l$ positive and negative
integers. The middle curve of Fig. \ref{fig2} shows the principal
peaks for $m=0,\pm 1,\pm 2$, and the combinations between the lowest
excited states ($l=\pm 1$), and the ground state for $m=\pm 1,\pm 2$.
Higher excited states contribute also because of the rational value
$\varepsilon/\omega=2.5$. Decreasing this ratio (top figure) the
number of peaks increase. The difference in their intensities is due
to the modulation of the corresponding Bessel function driven by
$\alpha$ (see later for a more detailed discussion).

In the presence of interaction ($g<1$) the discretization energy is
larger, $\varepsilon\propto 1/g$ (cf. eq.(\ref{dotenergy})), then for
a fixed frequency, increasing the interaction, we always end up into a
regime where only the principal values are observable.

\begin{figure}
\resizebox{0.48\textwidth}{!}{%
  \includegraphics{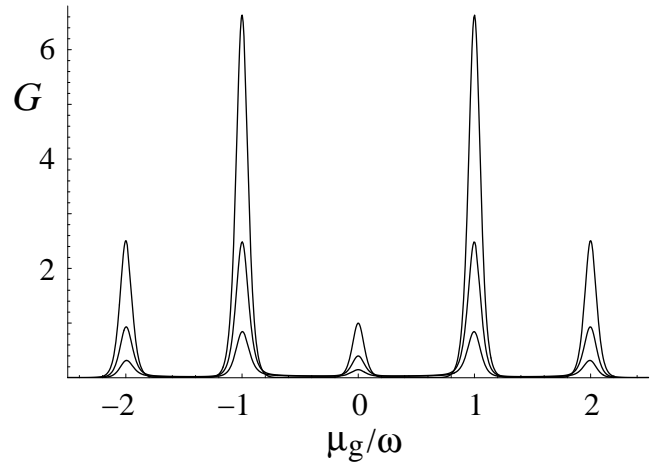}
}
\caption{Linear conductance, 
as a function of the gate energy $\mu_{g}/\omega$, for $\alpha=2$, 
$\varepsilon/\omega=5/g$, $k_{\rm B}T=\omega/30$, 
$\omega/\omega_{c}=10^{-4}$. The curves correspond to 
$g=1$,~$0.95,~0.9$ starting from the higher to the lower peak
intensities. The curves are normalized to the value of the central 
resonance ($\mu_{\rm g}=0$) for $g=1$.}
\label{fig3}       
\end{figure}

Figure \ref{fig3} shows the conductance for relatively low interaction
at fixed temperature, in the first regime $\varepsilon>\omega$.
One can easily see that while the positions of the principal sidebands
are not changed by the interactions, the intensities of the peak
resonances are drastically reduced.
Moreover, the stronger is the interaction the smaller is
the resolution of photo spectroscopy one can perform. For
$g<1$ the conductance assumes a smoother shape and the peaks become
progressively less distinguishable.
When lowering the temperature, the maxima of all of the resonance peaks scale
according to a power law similar to the dc-case \cite{braggio99}
\begin{equation}
\label{Gmax}
G_{\rm max}\propto(k_{\rm{B}}T)^{1/g-2}\,. 
\end{equation}
This expression shows the interplay of temperature and interaction and
strongly discriminates between two different regimes. In the {\it weak 
  interacting regime}, characterized by $1/2< g\le 1$, conductance
persists in showing a peak-like behavior which can be enhanced
decreasing temperature.  On the other hands in
the {\it strong interacting  regime} ($g\le 1/2$) the ac-field is no
longer able to split the dc-resonance into a series of peaks.
\begin{figure}
\resizebox{0.48\textwidth}{!}{%
  \includegraphics{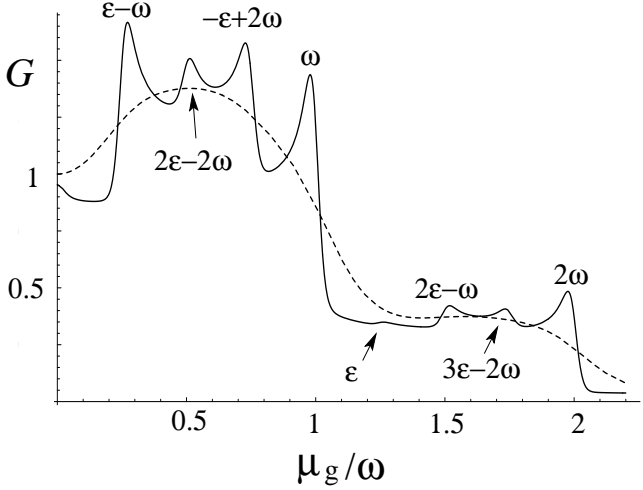}
}
\caption{Linear conductance 
as a function of $\mu_{\rm{g}}/\omega$, for $g=0.6$ and  
$k_{\rm B}T=\omega/15$ (dashed line), 
$k_{\rm B}T=\omega/100$ (full line) with  
$\alpha=2$, $\varepsilon/\omega=1.25$, and $\omega/\omega_{c}=10^{-4}$. 
The curves are normalized to the value of the central 
resonance ($\mu_{\rm g}=0$) at $k_{\rm B}T=\omega/15$.}
\label{fig4}       
\end{figure}
\begin{figure}
\resizebox{0.48\textwidth}{!}{%
  \includegraphics{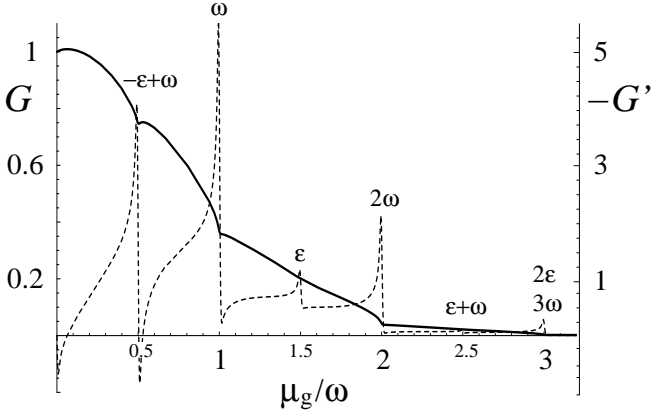}
}
\caption{Linear conductance
  (full line), and its derivative $G'=\omega \partial
  G/\partial\mu_{\rm g}$ (dashed line), in the strong interaction
  limit $g=0.4$, for $k_{\rm B}T=\omega/500$, $\alpha=2$,
  $\varepsilon/\omega=1.5$, and $\omega/\omega_{c}=10^{-4}$. The
  curves are normalized to the value of the conductance at the central
  resonance ($\mu_{\rm g}=0$).}
\label{fig5}       
\end{figure}
Figure \ref{fig4} shows the conductance as a function of $\mu_{\rm g}$
for weak interaction ($g=0.6$) and $\varepsilon\approx\omega$. The
dashed and full curves correspond to higher and lower temperatures,
respectively.  Decreasing temperature, the power law increas of the
peaks as in (\ref{Gmax}) allows to distinguish beween principal and
excited states. From the latter one can deduce the interaction
dependence of the plasmon modes.

Figure \ref{fig5} shows the conductance and its derivative for strong
interaction $g=0.4$.  Here the peaks are replaced by a step-like
behavior.  Their positions depend as above by the combination
$m\omega+l\varepsilon$.  In correspondence of each step the derivative
of the conductance presents sharp peaks.  Their line-shape strongly
depends on the interaction, with the maximum that scales in
temperature as $(\partial G/\partial\mu_{\rm g})\approx
(k_{\rm{B}}T)^{1/g-3}$.

The absence of resonance peaks in the photocon\-duc\-tan\-ce is a
behavior characteristic also of ``metallic'' quantum dots with a
continuum of electronic states (discretization energy much smaller
than the temperature)\cite{Kou1}. However, differently from our
discrete case where this absence is present only for strong
interactions $g\le 1/2$, here is always present at any interaction
$g\le 1$.  This is due to a sort of doubling of the interaction
between discrete and continuum dots \cite{Furusaki98,braggio99}.

\begin{figure}
\resizebox{0.48\textwidth}{!}{%
  \includegraphics{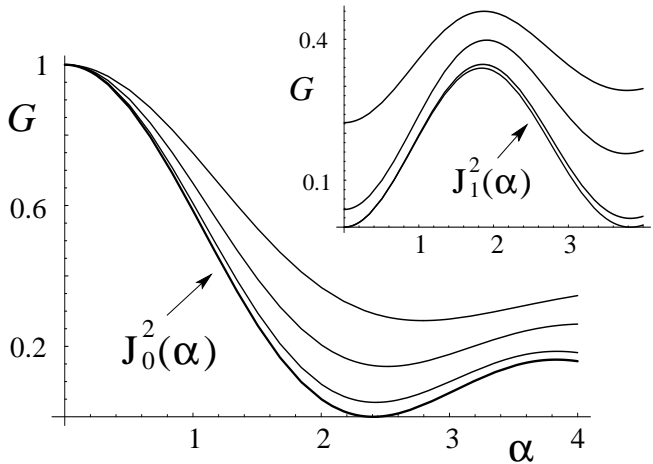}
}
\caption{Linear conductance, normalized to the
  height of the dc-resonance ($\alpha=0$),
  as a function of $\alpha=eV_{\rm ac}/\omega$, for $\mu_{\rm{g}}=0$,  
and $\mu_{\rm{g}}=\omega$ (inset), with
  $\varepsilon/\omega=4.5$, $g=0.8$, and $\omega/\omega_{c}=10^{-4}$.
Starting from the top to the bottom,
the different curves correspond to temperatures $k_{\rm
    B}T/\omega=1/2,1/5,1/20$ (main figure), and to $k_{\rm
    B}T/\omega=1/3,1/5,1/50$ (inset).  The
 curves indicated with arrows represent 
$J^2_0(\alpha)$ and $J^2_1(\alpha)$.}
\label{fig6}  
\end{figure}

Despite the positions and the shapes of the
photoconductance peaks are modified by the interaction, 
the corresponding height normalized to the dc value, 
still follows a universal relation.
For $k_{\rm B}T\ll\omega\ll\varepsilon$ and $g>1/2$, 
one can approximate the derivative of the rate (\ref{GAC}) 
at $\mu_{\rm g}=m\omega$ as
\begin{equation}
{\Gamma}'(m\omega)\approx {\Gamma}'(-m\omega)
\approx J^2_{m}(\alpha){\Gamma}'_{\rm dc}(0)\,,
\end{equation} 
where ${\Gamma}'_{\rm dc}(0)$ is the derivate of the dc-rate
(\ref{series}) for $\mu_{\rm g}=0$.
The photoconductance ratio assumes the universal form
\begin{equation}\label{andalf}
\frac{G(\mu_{\rm g}=m\omega)}{G_{\rm{dc}}(\mu_{\rm g}=0)}
\approx J^2_{m}(\alpha)\,.
\end{equation}
Figure \ref{fig6} represents the ratio (\ref{andalf}), in the
presence of interactions ($g= 0.8$), for $m=0$, and $m=1$ (inset) as a
function of $\alpha$, at different temperatures.  When the
temperature is sufficiently low the conductance ratio approaches the
Bessel function of the order of the corresponding principal sidebands.
Such a relation allows to select and enhance a given sideband at  the
cost of the others.  This can be achieved because the zeros of the
Bessel functions appear at different values of $\alpha$. 
The modulation of the peaks can then be controlled {\em
  independently} of the interaction simply by changing the ac-external
gate.

\section{Conclusions}
\label{sec:5}
In summary, we have evaluated the linear pho\-to-con\-duc\-tance of
a 1D quantum dot imbedded in a non-Fermi liquid.  We have considered
the sequential tunneling regime at temperatures smaller than the
discretization energy of the dot and the photon frequency.
Varying the frequency we treated both the ground state transport and
the transport via excited modes.  The latter are due to the collective
quantized plasmon modes in the dot, with an energy renormalized by the
interaction. We have shown that the presence of the interactions
dramatically changes the shape and the position of the exited peaks of the
photoconductance with respect to the non-interacting case. We found
two different transport regimes depending on the strength of the
interaction.  For weak interactions $1/2<g<1$ the linear conductance
presents localized peaks.  Their intensities scale at low temperatures
with a non-Fermi liquid power law driven by the interactions.  For
strong interactions $g\le 1/2$ the linear conductance keeps a compact
step-like profile and hardly discriminates between the different
sidebands. Peaks appear in this case in higher derivatives of the
conductance with respect to the gate voltage.

In evaluating the photoconductance we have considered
sequential tunneling processes in the limit of large barrier and high
frequencies. We have calculated the photo\-con\-duc\-tance
perturbatively in the tunneling matrix element $\Delta$. To be
consistent with this last assumption the conductance has to be much
smaller than $e^2/2\pi$, that of a clean wire .  Considering for example
the dc-conductance $G_{\rm dc}$ at the peak, this implies for the rate
(c.f. eq (\ref{dclincond}))
\begin{equation}
\label{vincolo}
\Gamma_{\rm dc}\ll k_{\rm B}T\,.
\end{equation}
As pointed already out by Furusaki \cite{Furusaki98}, this condition,
when $g<1/2$, once it is satisfied at some high temperature, remains 
valid until $T\to 0$. On the other hands at $g>1/2$ there is always a
critical temperature $k_{\rm B}T_{\rm
  c}=\varepsilon(\Delta/\varepsilon)^{g/(2g-1)}$ below which
perturbation expansion fails.  The ac-terms of the more general rate
$\Gamma^{(0)}$ follow similar argument being connected to $\Gamma_{\rm
  dc}$ via Bessel functions.  Note that because we always considered
temperatures smaller than the frequencies the relation (\ref{vincolo})
automatically satisfies the requirement of high frequencies $\omega\gg
\Gamma^{(0)}$.

The constraint (\ref{vincolo}) has to
be considered together with the condition that coherent processes,
like cotunneling, are negligible with respect to the sequential
ones.  Indeed at low enough temperatures one would expect that coherent
tunneling processes could dominate, especially in the region away
from resonance where sequential tunneling gives an almost negligible
contribution.  In \cite{Furusaki98} one can find an exhaustive discussion of
the interplay between cotunneling and sequential tunneling for
dc-transport. In particular it is shown that as long as (\ref{vincolo}) is
satisfied the sequential processes are the dominant ones near to the
peak resonance. Cotunneling contribution starts to become important 
in the tail of the peak. 

Another important point concerns the spin dynamics.  In the present
paper we considered a spinless electron gas, concentrating only on the
charge contribution. This was partly motivated by the spin-charge
separation effect present in a Luttinger liquid. However we expect
that the presence of the spin will introduce novel ac-effects.  In
particular the previous $g$ factor will be renormalized to an
effective $g_{\rm eff}$ given by $2/g_{\rm eff}=1/g +1/g_{\sigma}$.
The last term represents the contribution of the spin. For zero
exchange interactions $g_{\sigma}=1$. Moreover, we will have spin and
charge addition and discretization energies \cite{Tob}.  This will
enlarge the possible ac-transport channels involving more peaks in the
photoconductance due to the spin.  This interesting topics will be
the subject of future work.

We expect that the several above effects due to the interplay between
interaction and ac-fields can be observed in the future in the
transport through double barrier in cleaved-edge-over\-growth quantum
wires \cite{aus}, and in carbon nanotubes \cite{dekk}.

This  work has been supported by EU within TMR and RTN 
programmes, by Italian MURST via PRIN 2000.

\end{document}